\documentclass[conference,twoside]{IEEEtran}
\ifCLASSINFOpdf
\else
\fi

\usepackage{amssymb}
\usepackage{amsthm}
\usepackage[cmex10]{amsmath}
\DeclareMathOperator{\E}{\mathbb{E}}

\newcommand {\Define} {\stackrel {\Delta} {=}  }
\newcommand{\Eb}[1]{{ \mathbb{E}\left[ #1 \right] }}

\newcommand {\pu} {p_{\text{u}}}
\newcommand {\bw} {B_{\text{w}}}

\usepackage{bm}
\usepackage{mathtools}
\usepackage{fixltx2e}
\usepackage{epsfig,makeidx,color}
\usepackage{graphicx,dblfloatfix}
\usepackage{amsbsy}
\usepackage{amssymb}
\usepackage{euscript}
\usepackage{lipsum}
\usepackage[ruled,norelsize]{algorithm2e}
\usepackage{cite}
\usepackage{chngcntr}
\usepackage{lineno}
\usepackage[T1]{fontenc}
\usepackage{microtype}
\usepackage{wasysym}
\usepackage{footnote}

\newtheorem{remark}{\it Remark}

\usepackage[english]{babel}
    \usepackage[font=small,labelsep=space]{caption}
\usepackage{subcaption}    
\makeatletter
\def\citenoauxwrite#1{\begingroup
\@fileswfalse
\cite{#1}\relax
\endgroup}
\makeatother

\IEEEoverridecommandlockouts
\begin{document}
%
\title{Information Theoretic Performance of Periodogram-based CFO Estimation in Massive MU-MIMO Systems}
%
%
%
\author{\IEEEauthorblockN{Sudarshan Mukherjee and Saif Khan Mohammed}
\IEEEauthorblockA{ \thanks{The authors are with the Department of Electrical Engineering, Indian Institute of Technology Delhi (IITD), New delhi, India. Saif Khan Mohammed is also associated with Bharti School of Telecommunication Technology and Management (BSTTM), IIT Delhi. Email: saifkmohammed@gmail.com. This work is supported by EMR funding from the Science and Engineering
Research Board (SERB), Department of Science and Technology (DST),
Government of India.}}
}
\maketitle

\begin{abstract}
In this paper, we study the information theoretic performance of the modified time-reversal maximum ratio combining (TR-MRC) receiver (presented in \protect\citenoauxwrite{wcl2016}) with the spatially averaged periodogram-based carrier frequency offset (CFO) estimator (proposed in \protect\citenoauxwrite{gcom2016}) in multi-user massive MIMO systems. Our analysis shows that an $\mathcal{O}(\sqrt{M})$ array gain is achieved with this periodogram-based CFO estimator, which is same as the array gain achieved in the ideal/zero CFO scenario ($M$ is the number of base station antennas). Information theoretic performance comparison with the correlation-based CFO estimator for massive MIMO systems (proposed in \protect\citenoauxwrite{gcom2015}) reveals that this periodogram-based CFO estimator is more energy efficient in slowly time-varying channels.
\end{abstract}


%
\IEEEpeerreviewmaketitle

\vspace{-0.3 cm}
\section{Introduction}
%
%
%
%
Large scale antenna systems/massive multiple-input multiple-output  (MIMO) systems has been envisaged as one of the key technologies in the evolution of the next generation wireless communication systems \cite{Andrews, Boccardi}. In massive MIMO, the cellular base station (BS) is equipped with a large array of antennas (of the order of hundreds) to serve an unconventionally large number of single-antenna user terminals (UTs) simultaneously, in the same time-frequency resource \cite{Marzetta1}. Increasing the number of BS antennas opens up more available degrees of freedom, resulting in more effective suppression of multi-user interference (MUI) compared to the conventional single-antenna/small scale multi-antenna systems. It has been shown that for a given number of UTs, in a coherent multi-user massive MIMO system, with imperfect channel estimates, the required per-user transmit power in the uplink (to achieve a fixed desired per-user information rate) can be reduced as $\frac{1}{\sqrt{M}}$ with increasing $M$ (i.e. an $\mathcal{O}(\sqrt{M})$ array gain is achieved), where $M$ is the number of BS antennas \cite{Ngo1}.

\par However these existing results in massive MU-MIMO systems are based on the assumption of perfect frequency synchronization for coherent multi-user communication. In practice, carrier frequency offsets (CFOs) exist between the signals received at the BS from different UTs and the local oscillator at the BS. Existence of such CFOs, if unmitigated, would result in degradation of the information rate performance of the system. Although various techniques for frequency synchronization (CFO estimation/compensation) in small MIMO systems exist, it has been observed that those techniques are not amenable to practical implementation in massive MIMO systems, due to prohibitive increase in their complexity with increasing number of UTs \cite{Larsson2,gcom2015}.

In \cite{Larsson2}, the authors suggested an approximation to the joint maximum likelihood (ML) estimator for CFO estimation in multi-user (MU) massive MIMO systems. However, the CFO estimation technique presented in \cite{Larsson2}, requires multi-dimensional grid search and therefore has an exponential complexity with increasing number of UTs. Later in \cite{gcom2015}, a simple low-complexity (complexity independent of the number of UTs) correlation-based CFO estimator for massive MU-MIMO systems has been suggested. This CFO estimator however requires impulse-like pilots, which are highly susceptible to channel non-linearities (e.g. non-linear power efficient amplifier (PA) in the transmitters etc.), due to their high PAPR (peak-to-average-power ratio) characteristics. This problem of high PAPR pilots is later alleviated using a low-complexity (complexity linear with the number of UTs) spatially averaged periodogram-based CFO estimator proposed in \cite{gcom2016}, which uses low-PAPR constant envelope (CE) pilots. In \cite{gcom2016} it is shown that while the correlation-based CFO estimator in \cite{gcom2015} has less complexity, the periodogram-based CFO estimator proposed in \cite{gcom2016} is better in terms of the mean squared error (MSE) performance. 

\par However, while the information theoretic performance with the correlation-based CFO estimator has already been analyzed \cite{tvt2015,wcl2016}, no such result exists for the periodogram-based CFO estimator. Therefore, in this paper we derive the information theoretic performance with the periodogram-based CFO estimator, which also allows us to compare it to the information theoretic performance with the correlation-based CFO estimator.

\par \textsc{Contributions}: The novel results presented in this paper are summarized as follows: (i) firstly, we study the information rate performance of the modified time-reversal maximum ratio combining (TR-MRC) receiver proposed in \cite{wcl2016} with the periodogram-based CFO estimation for massive MU-MIMO uplink in the imperfect CSI scenario. Our study reveals that even with this new periodogram-based CFO estimator, an $\mathcal{O}(\sqrt{M})$ array gain is achievable (i.e. \textit{no loss in array gain performance compared to the ideal/zero CFO scenario}); (ii) 

a study of the trade-off between the information rate performance of the modified TR-MRC receiver (with the periodogram-based and correlation-based CFO estimators) versus the CFO estimation complexity reveals that the achievable information rate with the periodogram-based CFO estimator can be significantly better than that with the correlation-based CFO estimator at the cost of higher complexity; and (iii) further, it is also revealed that for slowly time-varying channels, the information rate performance with the periodogram-based CFO estimator is significantly better compared to that with the correlation-based CFO estimator, i.e., the periodogram-based CFO estimator is \textit{more energy efficient in slowly time-varying channels/low-mobility channels}. [\textbf{{Notations:}} $\E$ denotes the expectation operator and $(.)^{\ast}$ denotes the complex conjugate operator.]

 \vspace{-0.3 cm}

\section{System Model}
\begin{figure}[t]
 \centering
 \includegraphics[width= 3.4 in, height= 1.3 in]{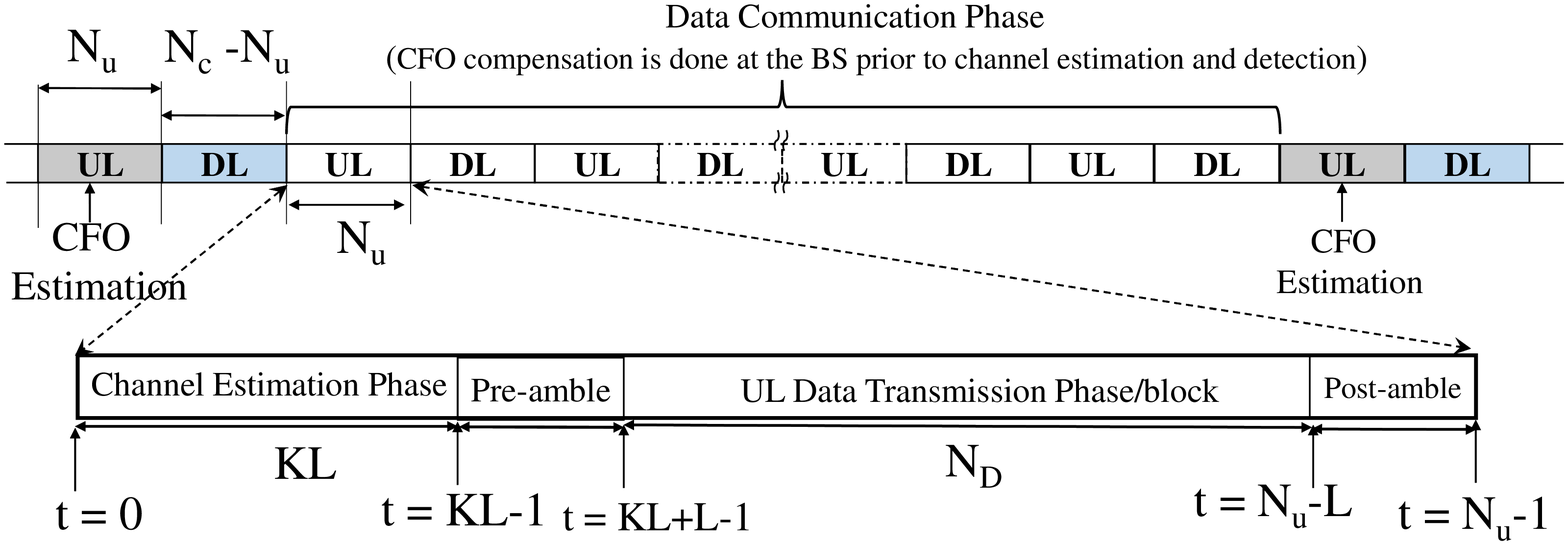}
 \caption {The communication strategy: CFO Estimation/Compensation and Data Communication. Here $N_c$ is the duration of coherence interval and the UL slot for data communication is $N_u$ channel uses.} 
 \label{fig:commstrat}
 \end{figure}

\vspace{-0.15 cm}
Let us consider a single-carrier single-cell massive MIMO BS, equipped with $M$ antennas, serving $K$ single antenna UTs simultaneously in the same time-frequency resource. Since a massive MIMO BS is expected to operate in time division duplexed (TDD) mode, i.e., each coherence interval is divided into an uplink (UL) slot, followed by a downlink (DL) slot. For coherent multi-user communication, frequency synchronization (i.e. CFO estimation/compensation) is important in massive MIMO systems. To this end, we consider a communication strategy, where the CFO estimation is performed at the BS in a special UL slot before data communication. In this slot, the UTs transmit special pilots to the BS. After CFO estimation, in the subsequent UL slots, at the BS, CFO compensation is performed, prior to channel estimation and UL receiver processing (see Fig.~\ref{fig:commstrat}). The special UL slot for CFO estimation is repeated every few coherence intervals, depending on how fast the CFOs change.

The CFO estimation/compensation technique presented in \cite{gcom2015,wcl2016} requires high PAPR impulse-like pilots, which are susceptible to PA non-linearities. Since massive MIMO systems are expected to be highly energy efficient, it is desired that we use low PAPR pilots signals for CFO estimation (to facilitate use of highly energy efficient non-linear PAs). The periodogram-based CFO estimation technique discussed in \cite{gcom2016} requires low PAPR constant envelope (CE) pilots. Specifically, for $K$ UTs, the $k^{\text{th}}$ UT would transmit a pilot $p_k[t] \, = \, e^{j\frac{2\pi}{K}(k-1)t}$, where $k = 1, 2, \ldots, K$ and $t = 0, 1, \ldots, N-1$. Here $N \leq N_u$ is the length of the pilot sequence and $N_u$ is the duration of the UL slot. Assuming the channel to be frequency-selective with $L$ memory taps, the pilot signal received at the $m^{\text{th}}$ BS antenna at time $t$ is given by

\vspace{-0.6 cm}

\begin{IEEEeqnarray}{rCl}
\label{eq:rxpilot}
\nonumber r_m[t] & = & \sqrt{\pu}\sum\limits_{q = 1}^{K}\sum\limits_{l = 0}^{L-1}h_{mq}[l]\, e^{j[\frac{2\pi}{K}(q - 1)(t - l) + \omega_q t]} \, + \, n_m[t]\\
& = & \sqrt{\pu} \sum\limits_{q=1}^{K} H_{mq} \, e^{j[\frac{2\pi}{K}(q - 1) + \omega_q]t} \, + \, n_m[t],
\IEEEeqnarraynumspace
\end{IEEEeqnarray}

\vspace{-0.3 cm}

\noindent where $H_{mq} \Define \sum\limits_{l = 0}^{L-1} h_{mq}[l] \, e^{-j \frac{2\pi}{K}(q-1)l}$ and $\omega_q$ is the CFO of the $q^{\text{th}}$ UT. Here $\pu$ is the average power transmitted by each UT in the uplink and $h_{mk}[l] \sim \mathcal{C}\mathcal{N}(0, \sigma_{hkl}^2)$ is the independent channel gain coefficient from the single-antenna of the $k$-th UT to the $m$-th antenna of the BS at the $l$-th channel tap. Also, $\{\sigma_{hkl} > 0\}, \, (l = 0, 1, \ldots, L-1; \, k = 1, 2, \ldots, K)$ is perfectly known at the BS and models the power delay profile (PDP) of the channel.

\vspace{-0.3 cm}

\subsection{Low-Complexity CFO Estimation Using Spatially Averaged Periodogram \cite{gcom2016}}

From \eqref{eq:rxpilot} it is clear that the signal received at the BS is simply a sum of complex sinusoids with additive Gaussian noise. Specifically, the frequency of the sinusoid received from the $k^{\text{th}}$ UT is $\frac{2\pi}{K}(k-1) + \omega_k$, where $\omega_k$ is the CFO of the $k^{\text{th}}$ UT. Intuitively, an estimate of this CFO of the $k^{\text{th}}$ UT, i.e., $\widehat{\omega}_k$ would be the difference between the frequency of the transmitted pilot (i.e. $\frac{2\pi}{K}(k-1)$) and the estimated frequency of the sinusoid received at the BS from the $k^{\text{th}}$ UT. An attractive low-complexity alternative to the joint ML frequency estimation is the periodogram technique \cite{Stoica}, which simply computes the periodogram of the received signal and chooses the $K$ largest peaks as the estimate of the $K$ frequencies. Since in massive MIMO systems, the required received power at the BS is expected to be small, spatial averaging of the periodogram computed at each of the $M$ BS antennas is performed \cite{gcom2016}.

\par We also assume that the CFOs from all UTs lie in the range $[-\Delta_{\text{max}}, \Delta_{\text{max}}]$ (where $\Delta_{\text{max}}$ is the maximum CFO for any UT). Therefore the frequency of the sinusoid received from the $k^{\text{th}}$ UT would lie in the interval $[\frac{2\pi}{K}(k - 1) - \Delta_{\text{max}}, \frac{2\pi}{K}(k - 1) + \Delta_{\text{max}}]$. For $\Delta_{\text{max}} < \frac{\pi}{K}$, these intervals for different UTs are non-overlapping,\footnote[1]{Note that the frequency range for two consecutive users (e.g. the $k^{\text{th}}$ UT and the $(k-1)^{\text{th}}$ UT) would be non-overlapping if and only if the maximum CFO $\Delta_{\text{max}}$ satisfy the following inequality $\big|\frac{2\pi}{K}(k-1) - \Delta_{\text{max}}\big| > \big|\frac{2\pi}{K}(k-2) + \Delta_{\text{max}}\big| \, \implies \, \big|\Delta_{\text{max}}\big| < \frac{\pi}{K}$. For a massive MIMO system with carrier frequency $f_c = 2$ GHz, communication bandwidth $\bw = 1$ MHz and maximum CFO of $0.1$ ($= \kappa$) PPM of $f_c$ \cite{Weiss}, the maximum CFO is given by $2\pi \kappa \, f_c/ \bw = \frac{\pi}{2500}$. Clearly with $K = 10$ and maximum delay spread of $5 \, \mu$s, i.e., $L = 5 \mu \text{s} \times \bw = 5$, we have $\Delta_{\text{max}} = \frac{\pi}{2500}\ll \frac{\pi}{KL} \ll \frac{\pi}{K} = \frac{\pi}{10}$, i.e., the frequency intervals of consecutive users are non-overlapping.} and therefore we need to compute the periodogram for the $k^{\text{th}}$ UT only in the interval $[\frac{2\pi}{K}(k - 1) - \Delta_{\text{max}}, \frac{2\pi}{K}(k - 1) + \Delta_{\text{max}}]$ over a fine grid of discrete frequencies. Thus the CFO estimate for the $k^{\text{th}}$ UT is given by\cite{gcom2016}

\vspace{-0.3 cm}

\begin{IEEEeqnarray}{rCl}
\label{eq:periodogram}
\widehat{\omega}_k = \arg \max\limits_{\theta \in \varXi} \, \underbrace{\overbrace{\frac{1}{M}\sum\limits_{m = 1}^{M}}^{\substack{\text{Spatial}\\ \text{averaging}}}\overbrace{\frac{1}{N}\Big | \sum\limits_{t = 0}^{N-1}r_m[t] \, e^{-j[\frac{2\pi}{K}(k - 1) + \theta]t}\Big |^2}^{\text{Periodogram computed at the $m^{\text{th}}$ BS antenna}}}_{ \Define \, \Phi_{k}(\theta)},
\IEEEeqnarraynumspace
\end{IEEEeqnarray}

\vspace{-0.3 cm}

\noindent where $\varXi \Define \{\varOmega(i) \Define \frac{2\pi}{N^{\alpha}}i \Big | |i| \leq T_0\}$, $T_0 \Define \lceil \frac{\Delta_{\text{max}}}{2\pi} N^{\alpha}\rceil$ and $\varOmega(i)$ denotes the discrete frequencies where the periodogram is computed. Note that the parameter $\alpha$ controls the resolution of the discrete frequencies in the set $\varXi$. Therefore it follows that with increasing $\alpha$ and fixed $N$, the MSE of CFO estimation, i.e., $\epsilon(\alpha) \Define \E[(\widehat{\omega}_k - \omega_k)^2]$ would decrease \cite{gcom2016}.

\section{Information Rate Analysis}
After the CFO estimation phase, the conventional data communication starts at $t = 0$ of the next UL slot (see Fig.~\ref{fig:commstrat}). The UTs transmit pilots for channel estimation sequentially in time for the first $KL$ channel uses. The UL data communication starts at $t = KL+L-1$ and continues for the next $N_D$ channel uses (i.e. from $t = KL+L-1$ till $t = KL+N_D+L-2$). The channel estimation phase and the UL data transmission phase are separated by a preamble sequence of $L-1$ channel uses.\footnote[2]{The symbols transmitted in the pre-amble and post-amble sequences (see Fig.~\ref{fig:commstrat}) are independent and identically distributed (i.i.d.) and are assumed to have the same distribution as the information symbols transmitted during the data communication phase, in order to ensure the correctness of the computed achievable information rate.} Since the duration of the UL slot is $N_u$ channel uses, it follows that $N_u = KL+ (L-1) + N_D + (L-1)$ and hence $N_D = N_u - KL - 2(L-1)$. For channel estimation, we assume that the $k^{\text{th}}$ UT transmits an impulse of amplitude $\sqrt{KL \pu}$ at $t = (k-1)L$ and zero elsewhere.\footnote[3]{Note that the use of impulse-like pilots for channel estimation is essential in order to have a fair comparison between the information rate achieved by the modified TR-MRC receiver with the periodogram-based CFO estimator (computed in this paper) and that with the correlation based CFO estimator (computed in \cite{wcl2016}).} Thus the received pilot at the $m^{\text{th}}$ BS antenna at time $t = (k-1)L+l$ is given by $r_m[(k - 1)L + l] = \sqrt{KL \pu}\, h_{mk}[l] \, e^{j \omega_k[(k - 1)L + l]} + n_m[(k - 1)L + l]$, where $m = 1, 2, \ldots, M$, $l = 0, 1, \ldots, L-1$ and $k = 1, 2, \ldots, K$. To estimate the channel gain coefficient, we first perform CFO compensation for the $k^{\text{th}}$ UT by multiplying $r_m[(k - 1)L + l]$ with $e^{-j\widehat{\omega}_k[(k - 1)L +l]}$ and then compute the ML channel estimate as $\widehat{h}_{mk}[l] \Define r_m[(k - 1)L +l]e^{-j\widehat{\omega}_k[(k-1)L + l]}/\sqrt{KL \pu} = \widetilde{h}_{mk}[l] + \frac{1}{\sqrt{KL \pu}}\widetilde{n}_m[(k-1)L + l]$. Here $\widetilde{n}_m[(k-1)L+l] \Define n_m[(k-1)L+l]e^{-j\widehat{\omega}_k[(k-1)L+l]} \sim \mathcal{C}\mathcal{N}(0, \sigma^2)$ and $\widetilde{h}_{mk}[l] \Define h_{mk}[l]e^{-j\Delta \omega_k [(k-1)L+l]} \sim \mathcal{C}\mathcal{N}(0,\sigma_{hkl}^2)$ is the effective channel gain coefficient and $\Delta \omega_k \Define \widehat{\omega}_k - \omega_k$ is the residual CFO after compensation.\footnote[4]{Both $h_{mk}[l]$ and $n_m[(k-1)L+l]$ have uniform phase distribution (i.e. circular symmetric) and are independent of each other. Clearly, rotating these random variables by fixed angles (for a given realization of CFOs and its estimates) would not change the distribution of their phases and they will remain independent. Therefore the distribution of $\widetilde{h}_{mk}[l]$ and $\widetilde{n}_{mk}[(k-1)L+l]$ would be same as that of $h_{mk}[l]$ and $n_m[(k-1)L+l]$ respectively.}

\begin{figure*}[!t]
\normalsize
\vspace*{-20pt}
\begin{IEEEeqnarray}{lCl}
\label{eq:xkhatmod}
\nonumber \widehat{x}_k[t] & = & \underbrace{M \, \sqrt{\pu}\Big(\sum\limits_{l=0}^{L-1}\sigma_{hkl}^2 \Big)\,\Eb{e^{-j\Delta\omega_k[t-(k-1)L]}} \, x_k[t]}_{= \, \text{ES}_k[t]} \, + \,  \underbrace{{\sqrt{\pu}\sum\limits_{m=1}^{M}\sum\limits_{l=0}^{L-1}\big|\widetilde{h}_{mk}[l]\big|^2\, e^{-j\Delta\omega_k[t-(k-1)L]}\, x_k[t] \, - \, \text{ES}_k[t]}}_{= \, \text{SIF}_k[t]} \, + \, \text{MUIN}_k[t]\\
& & \\
\label{eq:sinrkt}
\text{SINR}_k[t] & \Define & \dfrac{\Eb{|\text{ES}_k[t]|^2}}{\Eb{|\text{EW}_k[t]|^2}} =  \frac{\left(\Eb{e^{-j\Delta\omega_k[t - (k-1)L]}}\right)^2}{{\underbrace{\Bigg[1 - \left(\Eb{e^{-j\Delta\omega_k[t - (k-1)L]}}\right)^2\Bigg]}_{\Define \, \Eb{|\text{SIF}_k[t]|^2}} \, + \,\underbrace{\frac{1}{MK\gamma^2\theta_k^2}+ \frac{1}{M \gamma}\Bigg(1+\frac{1}{K\theta_k^2}\sum\limits_{q=1}^{K}\theta_q\Bigg)+\frac{1}{M\theta_k}\sum\limits_{q=1}^{K}\theta_q}_{\Define \,\, \Eb{|\text{MUIN}_k[t]|^2}} }}
\IEEEeqnarraynumspace
\end{IEEEeqnarray}
\hrulefill
\end{figure*}

\subsection{TR-MRC Receiver Processing}

After the channel estimation phase and the preamble transmission, the UL data communication starts. Let $x_k[t] \sim \mathcal{C}\mathcal{N}(0,1)$ be the i.i.d. information symbol transmitted by the $k^{\text{th}}$ UT at the $t^{\text{th}}$ channel use and $\pu$ be the average power transmitted by each UT. Therefore the received signal at the $m^{\text{th}}$ BS antenna at time $t$ is given by $r_m[t] = \sqrt{\pu}\sum_{q = 1}^{K} \sum_{l = 0}^{L - 1}h_{mq}[l]x_q[t - l]e^{j \omega_q t} + n_m[t]$, where $t = KL+L-1, \ldots, (N_D+KL+L-2)$. To detect $x_k[t]$, we use the modified TR-MRC receiver described in \cite{wcl2016}, i.e., we first perform CFO compensation for the $k^{\text{th}}$ UT and then pass the CFO compensated signal through the TR-MRC receiver. With TR-MRC processing of the CFO compensated signal, the detected information symbol for the $k^{\text{th}}$ UT at time $t$ is given by

\vspace{-0.7 cm}

{\begin{IEEEeqnarray}{rCl}
\label{eq:xkhat}
\nonumber \widehat{x}_k[t] & = & \sqrt{\pu}\sum\limits_{m=1}^{M} \sum\limits_{l = 0}^{L-1}\widehat{h}_{mk}^{\ast}[l]\underbrace{r_m[t+l]e^{-j\widehat{\omega}_k[t+l]}}_{\text{CFO Compensation}}\\
\nonumber & = & \sqrt{\pu}\bigg(\sum\limits_{m=1}^{M}\sum\limits_{l=0}^{L=1}\big|\widetilde{h}_{mk}[l]\big|^2\bigg) \, e^{-j\Delta\omega_k[t - (k-1)L]} \, x_k[t]\\ 
& & \,\,\,\,\,\,\,\,\,\,\,\,\,\,\,\,\,\,\,\,\,\,\,\,\,\,\,\,\,\,\,\,\,\,\,\,\,\,\,\,\,\,\,\,\,\,\,\, + \, \text{MUIN}_k[t] \, ,
\IEEEeqnarraynumspace
\end{IEEEeqnarray}}

\vspace{-0.6 cm}

where $\text{MUIN}_k[t]$ comprises of the inter-symbol interference (ISI), multi-user interference (MUI), channel estimation error and AWGN noise. In massive MIMO systems, it can be shown that with $M \to \infty$, the term $\sum_{m=1}^{M}\sum_{l=0}^{L-1}|\widetilde{h}_{mk}[l]|^2$ becomes almost deterministic\footnote[5]{As $M \to \infty$, the ratio of the standard deviation of $\sum\limits_{m=1}^{M}\sum\limits_{l=0}^{L-1}|\widetilde{h}_{mk}[l]|^2$ to its mean converges to zero.} due to channel hardening \cite{Marzetta1,Marzetta2}. Therefore an efficient communication strategy is to replace the effective channel gain component in the first term in the second line of R.H.S. of \eqref{eq:xkhat} by its mean value, i.e., $\text{ES}_k[t] \Define \sqrt{\pu}\,\Eb{\sum\limits_{m=1}^{M}\sum\limits_{l=0}^{L=1}\big|\widetilde{h}_{mk}[l]\big|^2 \, e^{-j\Delta\omega_k[t - (k-1)L]}}\, x_k[t]$ and create an additional term which would contain its variance around the mean, i.e., $\text{SIF}_k[t] \Define \Big(\sqrt{\pu}\,\sum\limits_{m=1}^{M}\sum\limits_{l=0}^{L=1}\big|\widetilde{h}_{mk}[l]\big|^2 \, e^{-j\Delta\omega_k[t - (k-1)L]}\, x_k[t] - \text{ES}_k[t]\Big)$.\footnote[6]{Here $\E[.]$ is taken across multiple coherence intervals and also across multiple CFO estimation phases.} Since CFO estimation is carried out in a separate special coherence interval, the residual CFO error is independent of the effective channel gain coefficient $\widetilde{h}_{mk}[l]$. Therefore we have $\text{ES}_k[t] = M \, \sqrt{\pu}\Big(\sum\limits_{l=0}^{L-1}\sigma_{hkl}^2 \Big)\,\Eb{e^{-j\Delta\omega_k[t-(k-1)L]}} \, x_k[t]$ and $\text{SIF}_k[t] = \sqrt{\pu}\Big(\sum\limits_{m=1}^{M}\sum\limits_{l=0}^{L-1}\big|\widetilde{h}_{mk}[l]\big|^2\, e^{-j\Delta\omega_k[t-(k-1)L]} \, - \, M\sum\limits_{l=0}^{L-1}\sigma_{hkl}^2 \, \Eb{e^{-j\Delta\omega_k[t-(k-1)L]}}\Big)  \, x_k[t]$. Thus from \eqref{eq:xkhat} we get \eqref{eq:xkhatmod}, where $\text{ES}_k[t]$ is treated as the useful signal term, and we relegate $\text{SIF}_k[t]$ to the effective interference and noise term $\text{EW}_k[t]$, i.e., $\text{EW}_k[t] \Define \text{SIF}_k[t] + \text{MUIN}_k[t]$. Hence from \eqref{eq:xkhat} we have $\widehat{x}_k[t] = \text{ES}_k[t] + \text{EW}_k[t]$. Note that the statistics of both $\text{ES}_k[t]$ and $\text{EW}_k[t]$ are functions of $t$. However for a given $t$, the realization of $\text{EW}_k[t]$ is i.i.d. across multiple UL data transmission blocks (i.e. coherence intervals). Therefore for each $t$, the effective channel between the $k^{\text{th}}$ UT and the BS reduces to a single-user SISO (single-input single-output) non-fading channel with additive noise, when viewed across multiple coherence intervals. Thus for $N_D$ channel uses, we would have $N_D$ SISO channels with distinct channel statistics. We therefore have separate codebooks, one for each of these $N_D$ channel uses. The data received in the $t^{\text{th}}$ channel use of every coherence interval is jointly decoded at the BS.\footnote[7]{This coding strategy has also been used in \cite{tvt2015,wcl2016,Phasenoise}.}

\vspace{-0.2 cm}

\subsection{Achievable Information Rate}

Since information symbol $x_k[t]$, residual CFO error $\Delta \omega_k = \widehat{\omega}_k - \omega_k$ and effective channel gain coefficient $\widetilde{h}_{mk}[l]$ are all independent random variables, it can be shown that $\Eb{\text{ES}_k[t] \text{EW}^{\ast}_k[t]} = 0$, i.e., the useful signal term is uncorrelated to the effective noise. Hence with Gaussian information symbols, the worst case uncorrelated noise (in terms of mutual information) would also be Gaussian with same mean and variance as $\text{EW}_k[t]$ \cite{Hasibi2}. Thus we have the following lower bound on the mutual information, i.e., $I(\widehat{x}_k[t]; x_k[t]) \geq \log_2\left(1 \, + \, \text{SINR}_k[t]\right)$, where $\text{SINR}_k[t]$ is defined in \eqref{eq:sinrkt} at the top of the page (note that $\theta_k \Define \sum_{l=0}^{L-1}\sigma_{hkl}^2$ and $\gamma = \frac{\pu}{\sigma^2}$ is the transmit SNR).\footnote[8]{This is due to the fact that $\Eb{\text{EW}_k[t]} = 0$, since $x_k[t]$ and the AWGN noise are both zero mean.} Therefore an achievable information rate for the $k^{\text{th}}$ UT is given by 

\vspace{-0.5 cm}

\begin{IEEEeqnarray}{rCl}
\label{eq:inforate}
I_k & = & \frac{1}{N_u}\sum_{t = KL+L-1}^{N_u - L}\log_2(1 + \text{SINR}_k[t]).
\IEEEeqnarraynumspace
\end{IEEEeqnarray}

\vspace{-0.4 cm}

\begin{figure}[t]
\vspace{-0.3 cm}
\centering
\includegraphics[width= 3.5 in, height= 2 in]{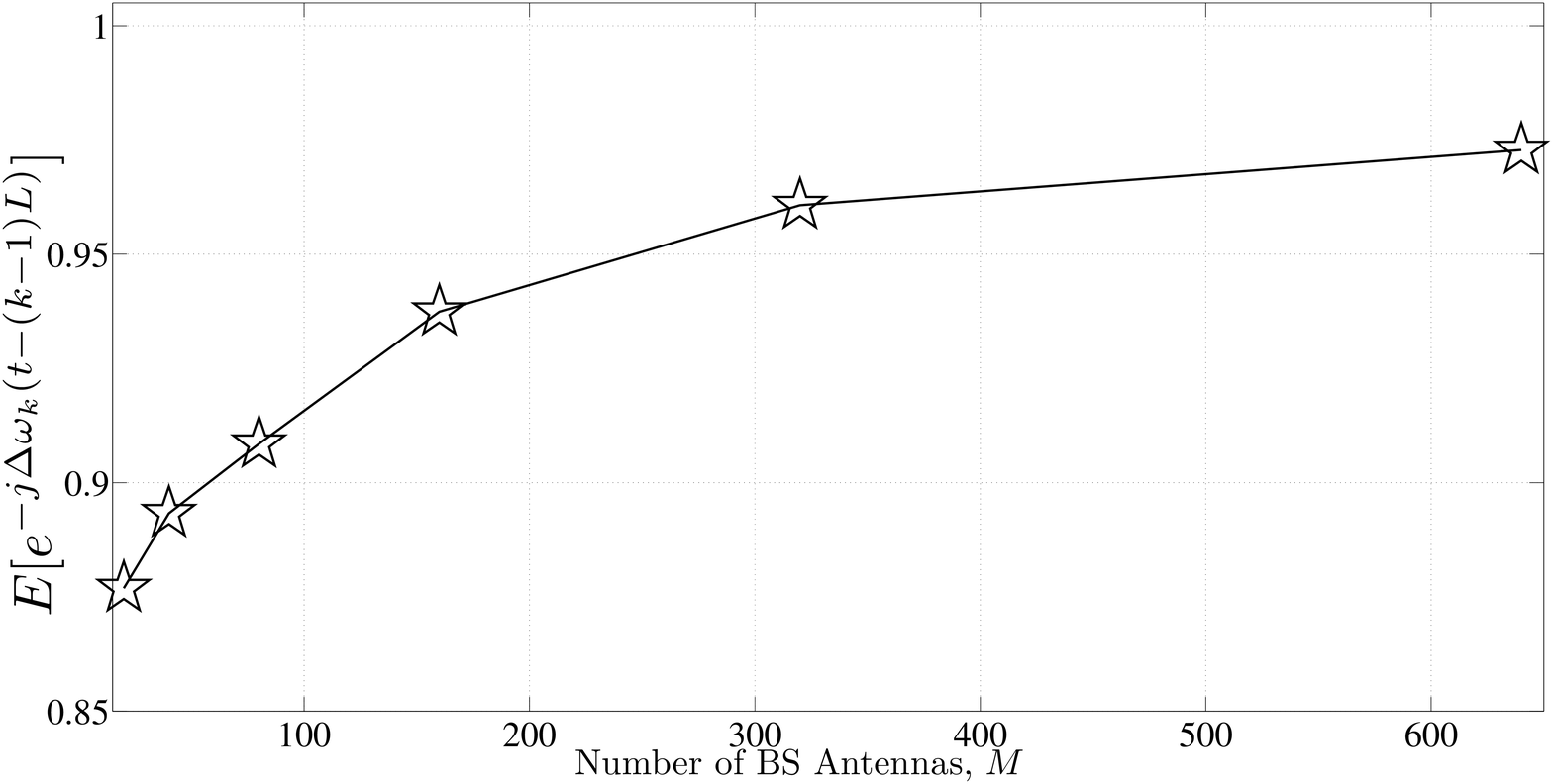}
\caption {Plot of the variation in $\Eb{e^{-j\Delta \omega_k(t - (k-1)L)}}$ with increasing number of BS antennas, $M$ for fixed $K = 10$, $N = 2000$, $N_u = 5000$, $L = 5$, $t = KL+N_D+L-2$ and $k = 1$, with the transmit SNR $\gamma$ decreasing $\propto \frac{1}{\sqrt{M}}$ with increasing $M$, starting at $\gamma = -14$ dB for $M = 20$.} 
\label{fig:expvarM}
\vspace{-0.3 cm}
\end{figure}

\begin{remark}
\label{arraygain}
(Achievable Array Gain)
\normalfont Analysis of the variances of various components of $\text{EW}_k[t]$ shows that $\Eb{\big|\text{EW}_k[t]\big|^2} = \Eb{\big|\text{SIF}_k[t]\big|^2} \, + \, \Eb{\big|\text{MUIN}_k[t]\big|^2}$ depends on the residual CFO error only through the variance of $\text{SIF}_k[t]$.\footnote[9]{It can be shown that $\Eb{\text{SIF}_k[t] \text{MUIN}_k^{\ast}[t]} = 0$, since $\Delta \omega_k$, $x_k[t]$ and $\widetilde{h}_{mk}[l]$ are all independent of each other.} We note that both the variances of $\text{ES}_k[t]$ and $\text{SIF}_k[t]$ depend on $\Eb{e^{-j\Delta\omega_k[t - (k-1)L]}}$, which in turn depends on the statistical distribution of ${(\widehat{\omega}_k - \omega_k)}$. From exhaustive numerical simulations, it can be easily shown that with the transmit SNR $\gamma \propto \frac{1}{\sqrt{M}}$, the term $\Eb{e^{-j\Delta\omega_k(t-(k-1)L)}}$ would converge to a constant value with increasing $M \to \infty$ (see Fig.~\ref{fig:expvarM}). Therefore the variances of $\text{ES}_k[t]$ and $\text{SIF}_k[t]$ would also converge to a fixed value with increasing $M \to \infty$ and $\gamma \propto \frac{1}{\sqrt{M}}$. Further from \eqref{eq:sinrkt}, we observe that the variance of $\text{MUIN}_k[t]$ converges to a non-zero positive constant value with increasing number of BS antennas $M \to \infty$ and $\gamma = \frac{\pu}{\sigma^2} \propto \frac{1}{\sqrt{M}}$.

\par Thus from the above discussion, it follows that $\text{SINR}_k[t]$ would converge to a constant as $M \to \infty$ with $\gamma \propto \frac{1}{\sqrt{M}}$. Hence from \eqref{eq:inforate} it can be concluded that the information rate of the modified TR-MRC receiver with the periodogram-based CFO estimation/compensation, converges to a constant with $\gamma \propto \frac{1}{\sqrt{M}}$ as $M \to \infty$. In other words, with a fixed desired information rate and fixed number of UTs, the required transmit SNR $\gamma$ would decrease roughly by $1.5$ dB with every doubling in the number of BS antennas $M$, i.e., an $\mathcal{O}(\sqrt{M})$ array gain is achieved (see the variation in the required transmit SNR $\gamma$ for $M = 320$ and $M = 640$ in Table~\ref{table:snrM}). This shows \textit{the interesting new result} that the periodogram-based CFO estimator does not degrade the achievable array gain in the residual CFO scenario (i.e., same as that for the correlation-based CFO estimator \cite{tvt2015,wcl2016}), when compared to the ideal/zero CFO scenario. \hfill \qed
\end{remark}

\vspace{-0.3 cm}

\section{Numerical Results and Discussions}

In this section, we study the information rate performance-complexity trade-off for the TR-MRC receiver with the periodogram-based CFO estimator. For Monte-Carlo simulations, we assume the following values for system parameters: carrier frequency $f_c = 2$ GHz, communication bandwidth $\bw = 1$ MHz and a maximum CFO equal to $0.1$ ($=\kappa$) PPM of $f_c$, i.e., $\Delta_{\text{max}} = 2\pi\kappa f_c/\bw = \frac{\pi}{2500}$. Also, we assume that the maximum delay spread of the channel is $T_d = 5 \mu$s, i.e., the number of channel memory taps $L = T_d\bw = 5$. At the start of every CFO estimation phase, the CFOs $\omega_k$ ($k = 1, 2, \ldots, K$) assume new values (independent of the previous ones), uniformly distributed in the interval $\big[-\frac{\pi}{2500}, \frac{\pi}{2500}\big]$. The PDPs are also assumed to be the same for all UTs and is given by $\sigma_{hkl}^2 = 1/L$, where $l = 0,1,\ldots, L-1$ and $k = 1, 2, \ldots, K$. The number of UTs are assumed to be $K = 10$.

\begin{savenotes}
\begin{table}[b]
\caption[position=top]{{\textsc{Minimum Required transmit SNR $\gamma = \frac{\pu}{\sigma^2}$ to achieve a fixed Per-User Information rate $I_k = 1$ bpcu, $K = 10$, $N = 2000$, $L = 5$ and UL slot duration $N_u = 5000$ channel uses.}}}
\label{table:snrM}
\centering
\begin{tabular}{| c | c | c | c | c | c |}
\hline
M & 40 & 80 & 160 & 320 & 640\\
\hline
\text{SNR} & -9.9 & -12.53 & -14.7 & -16.6 & -18.38 \\                 
\hline
\end{tabular}
\end{table}
\end{savenotes}

\par
\begin{figure}[t]
\centering
\includegraphics[width= 3.5 in, height= 2 in]{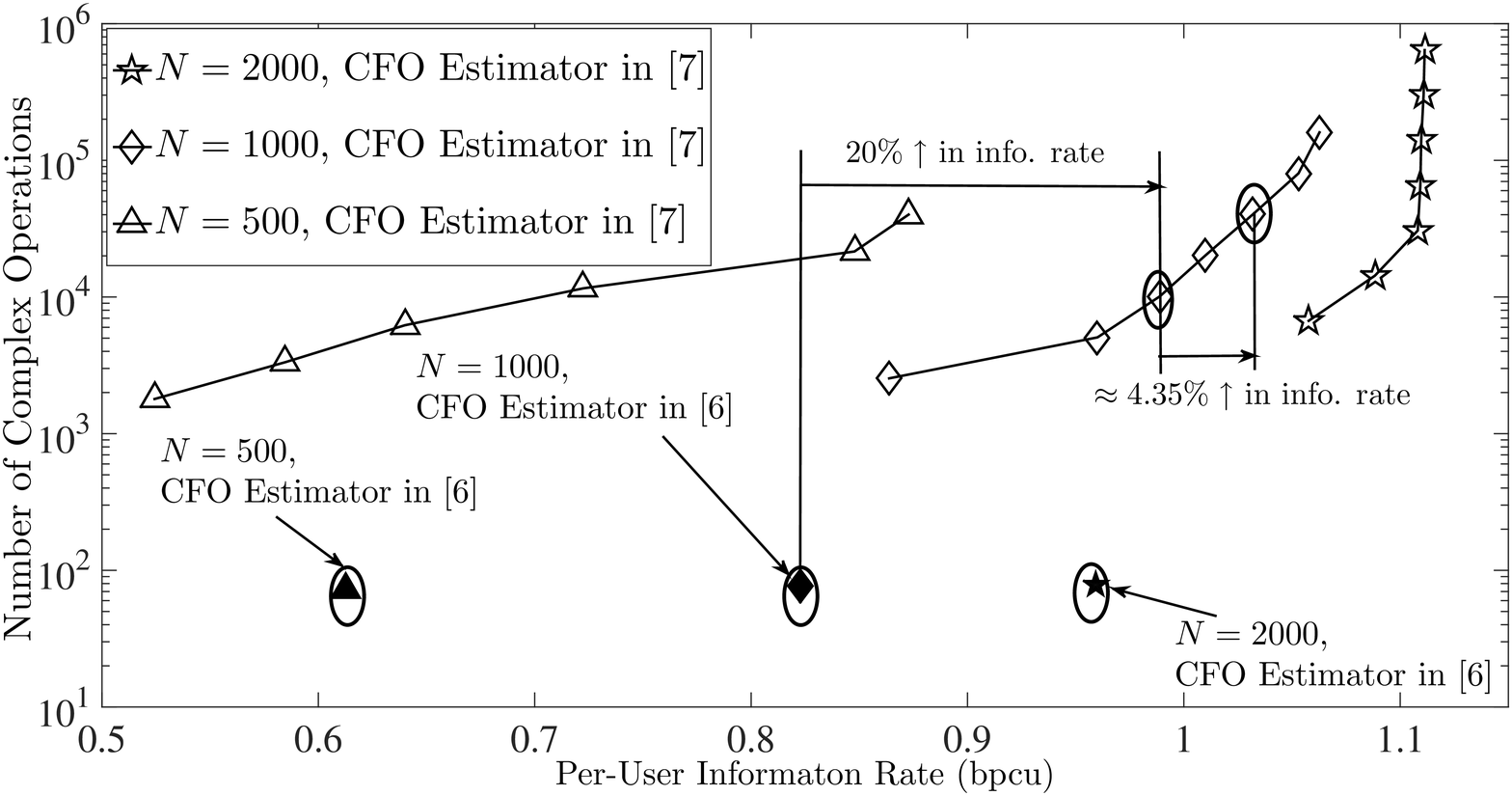}
\caption {Plot of the variation in the number of complex operations with increasing per-user information rate for fixed $K = 10$, $L = 5$, $SNR = -12$ dB, $M = 80$, $N_u = 5000$ and $N = 500, 1000$ and $2000$ respectively.} 
\label{fig:tradeinfo}
\vspace{-0.3 cm}
\end{figure}

 In Fig.~\ref{fig:tradeinfo}, we plot the variation in the total computational complexity (i.e. the number of complex floating point operations required for CFO estimation) versus the information rate of the $1^{\text{st}}$ UT for a fixed transmit SNR $\gamma = -12$ dB, fixed number of BS antennas $M = 80$, fixed duration of the UL slot $N_u = 5000$ and pilot length for CFO estimation $N = 500, 1000$ and $2000$ respectively.

The increase in the information rate (for a fixed transmit SNR $\gamma$, fixed $M$, $K$ and $N$) corresponds to the increase in $\alpha$, since with increasing $\alpha$ the resolution of the CFO estimation in \eqref{eq:periodogram} increases, thereby reducing the MSE of CFO estimation. Reduction in the MSE of CFO estimation reduces the variance of the SIF term in the denominator in \eqref{eq:sinrkt} and this leads to an increase in the information rate. However, the information rate does not increase unboundedly with increasing $\alpha$ and is seen to saturate for values of $\alpha$ beyond a critical value. This is observed in Fig.~\ref{fig:tradeinfo}, where for $N = 1000$, the increase in the information rate is only $\approx 4.35 \%$, when $\alpha$ is increased from $1.6$ to $1.8$, though the complexity increases rapidly by a factor of approximately $4$ (see Fig.~\ref{fig:tradeinfo}). Therefore from the point of view of the complexity-performance trade-off, it appears that it is optimal to operate with $\alpha$ equal to this critical value. In this paper, we have defined this critical value $\alpha^{\star}$ as the smallest possible value of $\alpha$ for a fixed transmit SNR $\gamma$, $M$, $K$ and $N$, such that $|(I_k(\alpha) - I_k(\alpha + \Delta \alpha))/I_k(\alpha)| < \delta$, for a given $\delta = 0.02$ and $\Delta \alpha = 0.1$.

\par In Fig.~\ref{fig:tradeinfo}, we also plot the required number of complex operations for CFO estimation using the correlation-based CFO estimator proposed in \cite{gcom2015}. Note that while this CFO estimator is indeed superior to the periodogram-based CFO estimator proposed in \cite{gcom2016} in terms of complexity, the information rate performance of the TR-MRC receiver with the periodogram-based CFO estimator is comparatively better than that of the TR-MRC receiver with correlation-based CFO estimator when $\alpha$ is sufficiently large. For instance, with $N = 1000$ and $\alpha = \alpha^{\star} = 1.6$, the information rate with the periodogram-based CFO estimator is approximately $20\%$ more than that of the correlation-based CFO estimator (see Fig.~\ref{fig:tradeinfo}).

\par
\begin{figure}[t]
\centering
\includegraphics[width= 3.5 in, height= 2 in]{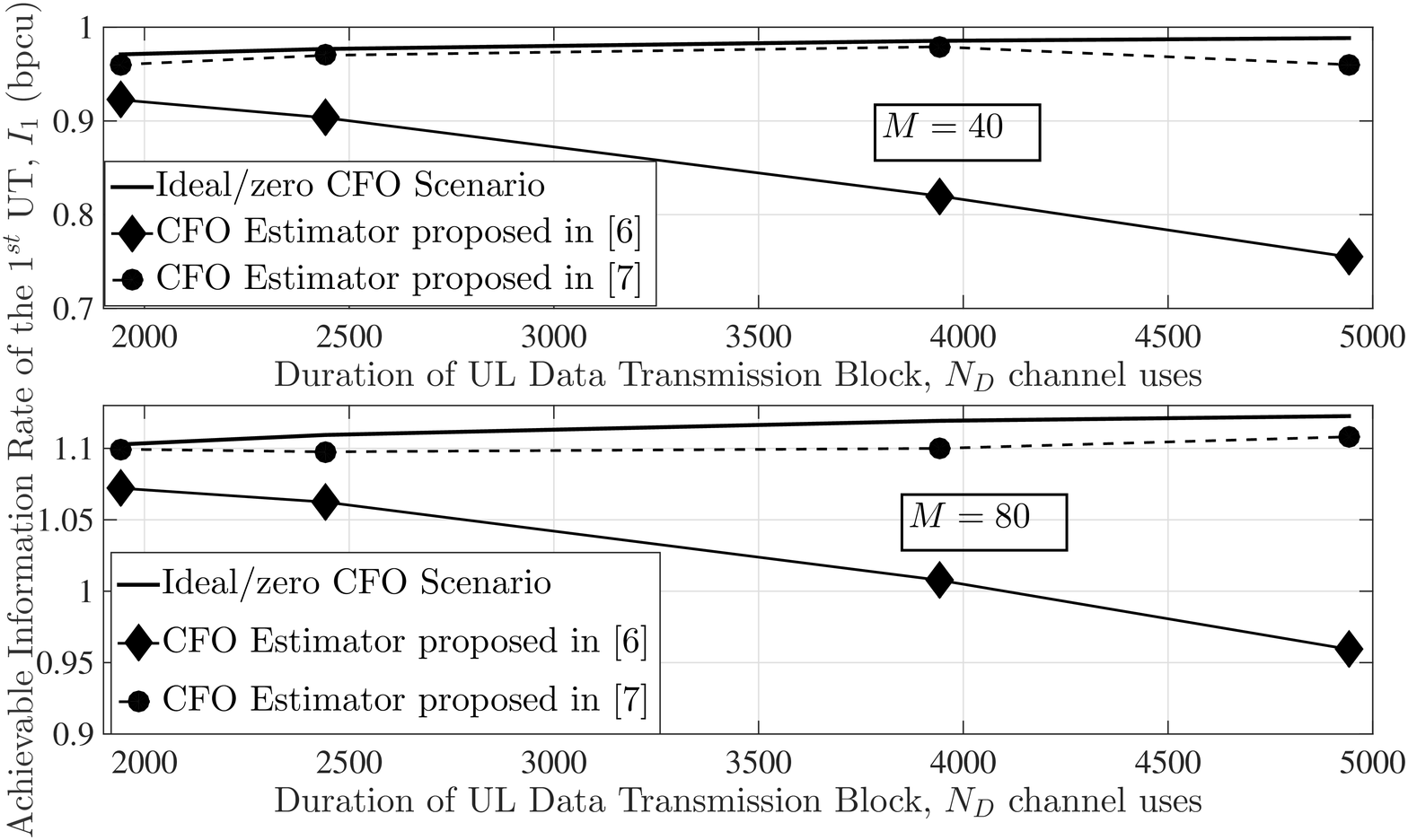}
\caption {Plot of the variation of achievable per-user information rate versus duration of UL data transmission block, $N_D$ channel uses with fixed $N = 2000$, $K = 10$, $L = 5$, $SNR = -10$ dB for $M = 40$ and $SNR = -12$ dB for $M = 80$.} 
\label{fig:infoblock}
\vspace{-0.3 cm}
\end{figure}

 Subsequently, in Fig.~\ref{fig:infoblock} we plot the variation in the achievable information rate for the first user with increasing duration of UL data transmission block ($N_D$), for a fixed transmit SNR $\gamma = -10$ dB for $M = 40$ antennas and fixed SNR $\gamma = -12$ dB for $M = 80$ antennas. We consider the following three scenarios: (a) the ideal/zero CFO scenario (solid line with no marker); (b) the residual CFO scenario with the periodogram-based CFO estimator in \cite{gcom2016} (dashed line with filled circles); and (c) the residual CFO scenario with the correlation-based CFO estimator in \cite{gcom2015} (solid line with filled diamonds). For the residual CFO scenario with the periodogram-based CFO estimator, we set $\alpha$ to its critical value $\alpha^{\star}$ (defined in the discussion for Fig.~\ref{fig:tradeinfo}). Note that in each scenario, the information rate initially increases with increasing $N_D$, due to the increase in the fraction of UL slot used for data transmission. However for the residual CFO scenarios, with further increase in $N_D$, the information rate starts to decrease. This decrease is due to the fact that the channel estimates acquired at the beginning of UL slot become stale (i.e. the accumulated phase error due to residual CFO becomes significantly large with increasing time-lag between the channel estimation phase and the time instances when the information symbols are received). Interestingly, it is observed that with increasing $N_D$, this performance degradation is much more pronounced with the correlation-based CFO estimator, compared to the periodogram-based CFO estimator. For instance when $N_u = 2000$ (i.e. $N_D = N_u - KL-2(L-1) = 1942$) and $M = 40$, the loss in the information rate w.r.t. the ideal/zero CFO scenario is $1.12\%$ and $5\%$ for the periodogram-based CFO estimator and the correlation-based CFO estimator respectively. However, when $N_D$ is increased to $4942$ (i.e. $N_u = 5000$), this loss in the information rate performance increases to $2.87\%$ and $23.62\%$ for the periodogram-based CFO estimator and the correlation-based CFO estimator respectively. It is therefore observed that the periodogram-based CFO estimator is more robust to the residual CFO when compared to the correlation-based CFO estimator. Similar observations can be made from Fig.~\ref{fig:infoblock} when $M=80$.

\par
\begin{figure}[t]
\centering
\includegraphics[width= 3.5 in, height= 2 in]{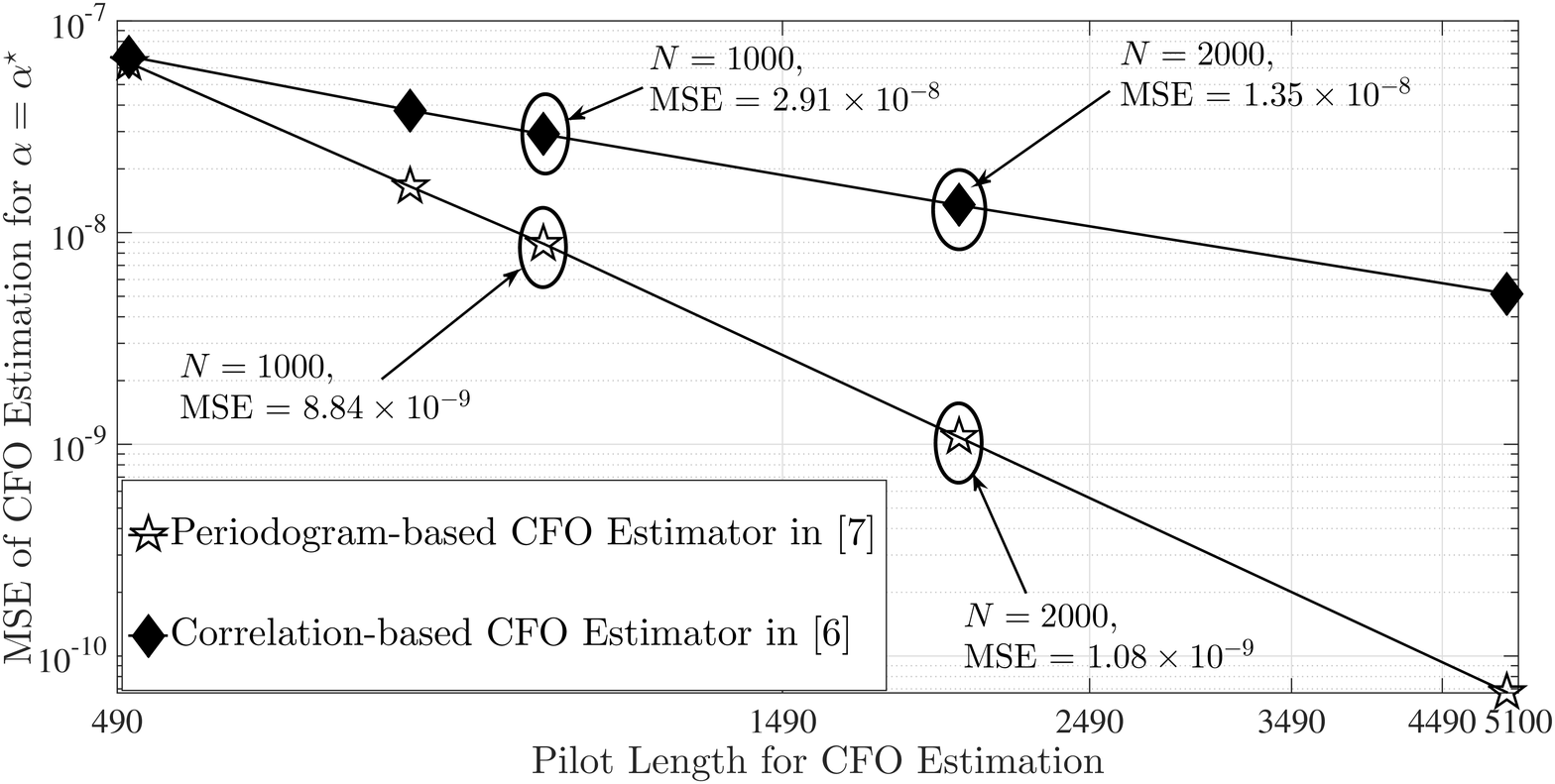}
\caption {Plot of the variation in the MSE ($\alpha = \alpha^{\star}$) with increasing pilot length for CFO Estimation $N$ for the periodogram-based CFO Estimator. Fixed Parameters: $K = 10$, $M = 80$, $L = 5$, transmit SNR $\gamma = -10$ dB.} 
\label{fig:msevsN}
\vspace{-0.3 cm}
\end{figure}

 This robustness of the periodogram-based CFO estimator is due to the fact that for a fixed transmit SNR $\gamma$, $M$ and $K$, the MSE of the correlation based CFO estimator is $\propto 1/N$ (see equation (1) in \cite{gcom2015}), while the MSE of the periodogram based CFO estimator is proportional to $1/N^{3}$ for sufficiently large $N$, as can be verified from Fig.~\ref{fig:msevsN}.\footnote[10]{In Fig.~\ref{fig:msevsN} we have plotted the MSE ($\alpha = \alpha^{\star}$) as a function of the pilot length $N$, on a log-log scale for fixed $M = 80$, $K = 10$, $L = 5$ and transmit SNR $\gamma = -10$ dB. It is observed that the slope of the curve is approximately $-3$ for sufficiently large $N$.} This shows that when $N_D$ is large, i.e., in slowly time-varying channels/low-mobility channels, the periodogram-based CFO estimator ($\alpha = \alpha^{\star}$) can yield a better information rate performance than the correlation-based CFO estimator for the same transmit power. Hence, the periodogram-based CFO estimator is expected to be more energy efficient in slowly time-varying channels.

%

%


\ifCLASSOPTIONcaptionsoff
  \newpage
\fi



%


\bibliographystyle{IEEEtran}
\bibliography{IEEEabrvn,mybibn}

\nocite{gcom2015,gcom2016,wcl2016}

\end{document}